\documentclass[preprint,showpacs,aps]{revtex4}
\usepackage{epsfig}
\usepackage{graphicx}

\newcommand{\be}{\begin{equation}}
\newcommand{\ee}{\end{equation}}
\newcommand{\ba}{\begin{eqnarray}}
\newcommand{\ea}{\end{eqnarray}}

\begin{document}

\title{Fine structure of the Poynting-Robertson effect \\
       for a luminous spinning relativistic star}

\author{Jae Sok Oh\footnote{e-mail: ojs@astro.snu.ac.kr}}

\affiliation{Department of Physics and Astronomy, FPRD, Seoul
National University, Seoul 151-742, Korea}

\author{Hongsu Kim\footnote{e-mail: hongsu@kasi.re.kr}}

\affiliation{KVN, Korea Astronomy
and Space Science Institute, Daejeon 305-348, KOREA}

\author{Hyung Mok Lee\footnote{e-mail: hmlee@astro.snu.ac.kr}}

\affiliation{Department of Physics and Astronomy, FPRD, Seoul
National University, Seoul 151-742, Korea.}

\begin{abstract}

As a sequel to our recent works challenging toward the systematic
inclusion of the effect of radiation on the trajectory of a test
particle orbiting around a luminous spinning relativistic star eventually
aiming at its application to the accretion flow. We explore in the
present work the fine structure of the trajectory of test particle
just entering the ``suspension orbit" under the purpose of a detailed investigation
of test particle's trajectory in the vicinity of the ``suspension orbit".
We end up with a rather puzzling behavior that, contrary to our expectation,
the specific angular momentum of the test particle instantly rises instead of decreasing
monotonically just before the test particle enters the ``suspension orbit".

\end{abstract}


\maketitle

\newpage
\begin{center}
{\rm\bf I. INTRODUCTION}
\end{center}

Astrophysical accretion flow onto massive or compact stars is one of
the major concerns in Astronomy and Astrophysics. In the current
treatment of the accretion process, the effect of radiation 
on the inflow has been poorly addressed. Therefore in our recent
works(\cite{OKL2010}, \cite{OKL2011}), we have been challenging
toward the systematic inclusion of the effects of radiation in the
accretion process. To this end, we explored the effect of the
radiation from a central star on the motion of a single test
particle when the central star has the angular momentum and
a finite radius to realize that there exists the ``suspension orbit"
that corresponds to the "critical point" in \cite{AEL90}. There
(\cite{OKL2010}), the ``suspension orbit" has been discovered for the first time
and it can be defined as an orbit in which the test particle hovers
around the central star at uniform velocity (for more detail, see
\cite{OKL2010}).

In the present work, we would like to pursue this effort further and
in this sense, the present work can be regarded as a sequel to our
earlier works (\cite{OKL2010}, \cite{OKL2011}). To be more specific,
we explore the fine structure of the trajectory of test particle
just entering the ``suspension orbit" under the purpose of a
detailed investigation of test particle's trajectory in the vicinity
of the ``suspension orbit". To summarize the main result of our
present work, as we shall see shortly in the text, we encounter a
rather puzzling behavior that, contrary to our expectation, the
specific angular momentum of the test particle instantly rises instead of
decreasing monotonically just before the test particle enters the
``suspension orbit".  Indeed, we find it anomalous as it contradicts
to the Kepler's law which is obviously the first principle. In the
text of this paper, we will attempt to address the relevant physical
interpretation of it.

\begin{center}
{\rm\bf II. FINE STRUCTURE OF THE TRAJECTORY NEAR THE ``SUSPENSION ORBIT"}
\end{center}

As we mentioned in the introduction above, in this section we now
would like to report on the fine structure of the test particle's
trajectory near the ``suspension orbit" which exhibits some
interesting features. To summarize the motivation behind our current
research, so far we have been exploring the dynamics of a test
particle orbiting around a luminous relativistic stars
(\cite{OKL2010} and \cite{OKL2011}). And the purpose of such study
is to have some insight into the behavior of the accretion flow onto
the relativistic stars emitting radiation such as the AGN or the
X-ray binaries. In other words, we would like to understand the
effect of radiation on the accretion inflow toward stars with strong
gravity which has not been addressed in a systematic manner in the
literature. To be more specific, based upon our earlier discovery of
the ``suspension orbit" (\cite{OKL2010}) which can be thought of as
the generalization of the critical point (\cite{AEL90}) for the case
of sufficiently luminous non-rotating relativistic stars and the
detailed study of the advent and the effect of the counter drag
forces in our recent works (\cite{OKL2010} and \cite{OKL2011}), in
the present work, we studied the behavior of the trajectory of the
test particle just entering the ``suspension orbit" that we have
discovered in our earlier study (\cite{OKL2010}). This exploration
can be thought of as a detailed investigation of test particle's
trajectory in the vicinity of the ``suspension orbit".  We hope such
a study of the fine structure would help our understanding of the
nature of ``suspension orbit" (\cite{OKL2010}). As can be seen in a
moment, we will consider the co-rotating since we are interested in the dynamics
of a test particle approaching a sufficiently luminous spinning
relativistic star.  

Now, the geodesic equation for the azimuthal component of test particle's velocity is given below:
\begin{eqnarray}
   \frac{dU_{\phi}}{d\tau}
      &=& - \frac{2}{(1+\cos\alpha)}
            \left(1-\frac{2M}{r}\right)^{-2}
            \left(\frac{M}{r^2}\right)
            \left(\frac{L^{\infty}}{L^{\infty}_{Edd}}\right)
            U^2_{t}U_{\phi}
            \nonumber   \\
      &-&  2\left(1-\frac{2M}{r}\right)^{-1}
            \left(\frac{M}{r^2}\right)
            \left(\frac{L^{\infty}}{L^{\infty}_{Edd}}\right)
            U_{t} U_{r} U_{\phi}
            \nonumber   \\
      &-&   \frac{2(1+\cos\alpha+\cos^2\alpha)}{3(1+\cos\alpha)}
            \left(\frac{M}{r^2}\right)
            \left(\frac{L^{\infty}}{L^{\infty}_{Edd}}\right)
            U^2_{r}U_{\phi}
            \nonumber   \\
      &-&   \frac{(2-\cos\alpha-\cos^2\alpha)}{3(1+\cos\alpha)}
            \left(1-\frac{2M}{r}\right)^{-1}
            \left(\frac{M}{r^2}\right)
            \left(\frac{L^{\infty}}{L^{\infty}_{Edd}}\right)
            \left(\frac{r^2+U^2_{\phi}}{r^2}\right)U_{\phi}
            \nonumber   \\
      &-&   \frac{4}{(1+\cos\alpha)}\omega(r)
            \left(1-\frac{2M}{r}\right)^{-2}
            \left(\frac{M}{r^2}\right)
            \left(\frac{L^{\infty}}{L^{\infty}_{Edd}}\right)
            U_{t}U^2_{\phi}
            \nonumber   \\
      &-&   2\omega(r)
            \left(1-\frac{2M}{r}\right)^{-1}
            \left(\frac{M}{r^2}\right)
            \left(\frac{L^{\infty}}{L^{\infty}_{Edd}}\right)
            U_{r}U^2_{\phi}
            \nonumber   \\
      &-&   \frac{2(2-\cos\alpha-\cos^2\alpha)}{3(1+\cos\alpha)}
            \left(\frac{\mathcal{J}(r)}{r}\right)
            \left(1-\frac{2M}{r}\right)^{-2}
            \left(\frac{M}{r^2}\right)
            \left(\frac{L^{\infty}}{L^{\infty}_{Edd}}\right)
            U_{t}U^2_{\phi}
            \nonumber   \\
      &-&   \frac{(2-\cos\alpha-\cos^2\alpha)}{3(1+\cos\alpha)}
            \left(r\mathcal{J}(r)\right)
            \left(1-\frac{2M}{r}\right)^{-2}
            \left(\frac{M}{r^2}\right)
            \left(\frac{L^{\infty}}{L^{\infty}_{Edd}}\right)
            U_{t}
            \nonumber   \\
      &-&   \frac{\sin^2\alpha}{4}
            \left(r\mathcal{J}(r)\right)
            \left(1-\frac{2M}{r}\right)^{-1}
            \left(\frac{M}{r^2}\right)
            \left(\frac{L^{\infty}}{L^{\infty}_{Edd}}\right)
            \left(\frac{r^2+2U^2_{\phi}}{r^2}\right)U_{r},
\end{eqnarray}
where $\sin\alpha = \left(\frac{R}{r} \right){\left(\frac{1 -
2M/r}{1 - 2M/R}\right)}^{1/2}$ (see \cite{AEL90}) for the radius of
the star $R \geq 3M$, the Eddington luminosity $L^{\infty}_{Edd}
\equiv 4\pi m M/\sigma$ is the luminosity of a spherically symmetric
source such that at infinity the outward radiation force  balances
the inward gravity (see \cite{LM95}), $L^{\infty}$ is the luminosity
of the star as measured by an observer at infinity, and
$\mathcal{J}(r)$ is given by
\begin{eqnarray}
   \mathcal{J}(r) = 8j\left(\frac{r}{M}\right)
                    \left(\frac{M^3}{R^3}-\frac{M^3}{r^3}\right)
                + 4v\left(\frac{r}{R}\right)(1-2M/R)^{1/2},
                \nonumber
\end{eqnarray}
where $j \equiv
cJ/(GM^2)$ is the dimensionless angular momentum of the star 
and the average azimuthal velocity $v$ of the radiation
source as measured by an observer in the LNRF (Locally Non-Rotating
Frame; see \cite{B70}; \cite{BPT72}) is calculated to be
\begin{eqnarray}
      v = \frac{1}{\pi} j \left(\frac{M^2}{R^2}\right)
          \left(1-\frac{2M}{R}\right)^{-1/2}
          \left[5\left(\frac{R}{M}\right)-4\right].
          \nonumber
\end{eqnarray}

We now turn to the numerical solutions of the test particle's
specific angular momentum that can be represented by the plots
(Fig.1). Interestingly enough, it turns out that the
numerical solutions reveal contrasting characteristic features
between the case when the ``suspension orbit" develops just above or
below the surface of the star and the other case when the
``suspension orbit" emerges at a distance from the star. Therefore,
we now start with the first case when the ``suspension orbit"
develops just above or below the surface of the star.

\begin{figure}
\centering
\includegraphics[width=0.4\textwidth]{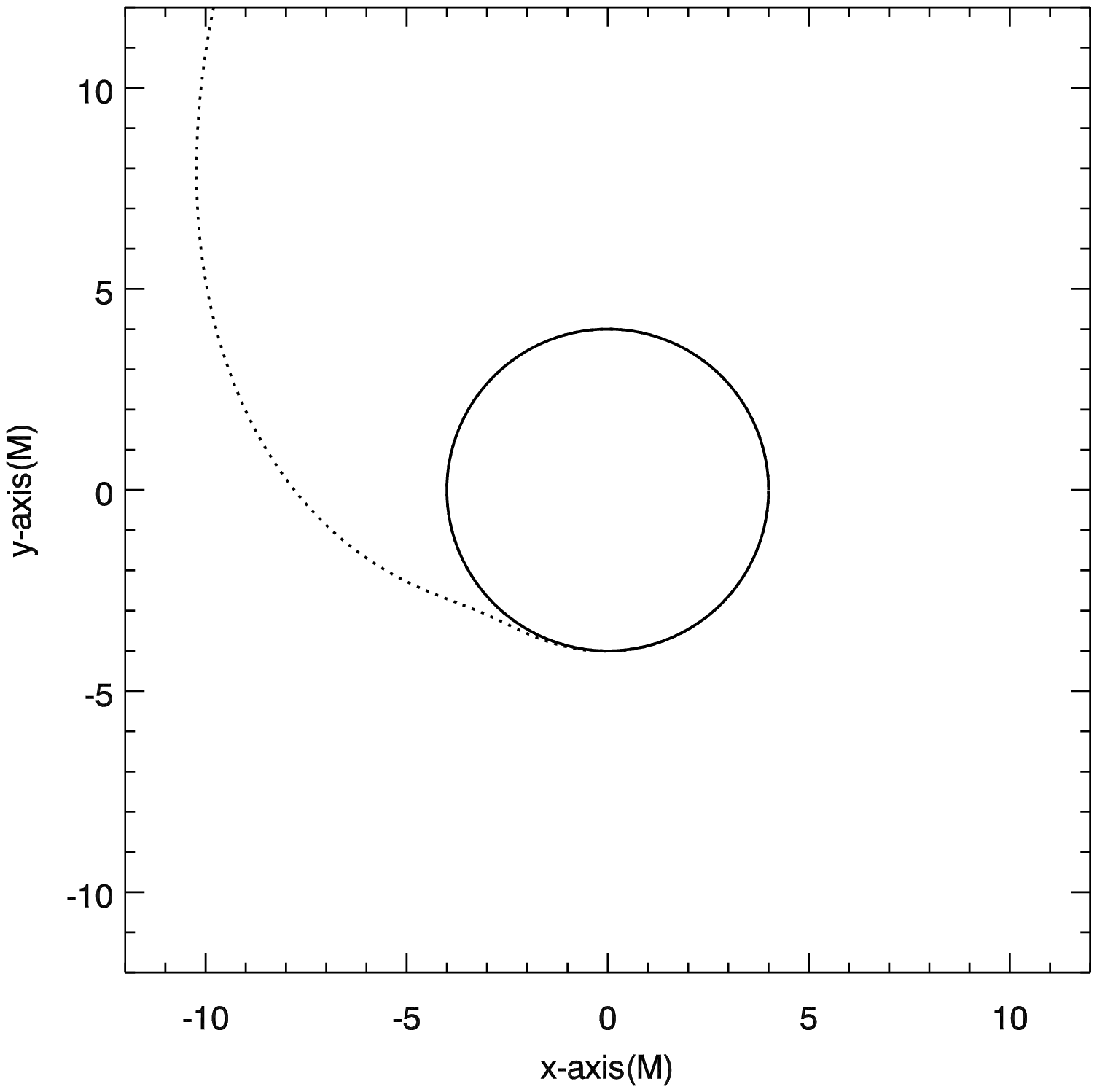}
\includegraphics[width=0.4\textwidth]{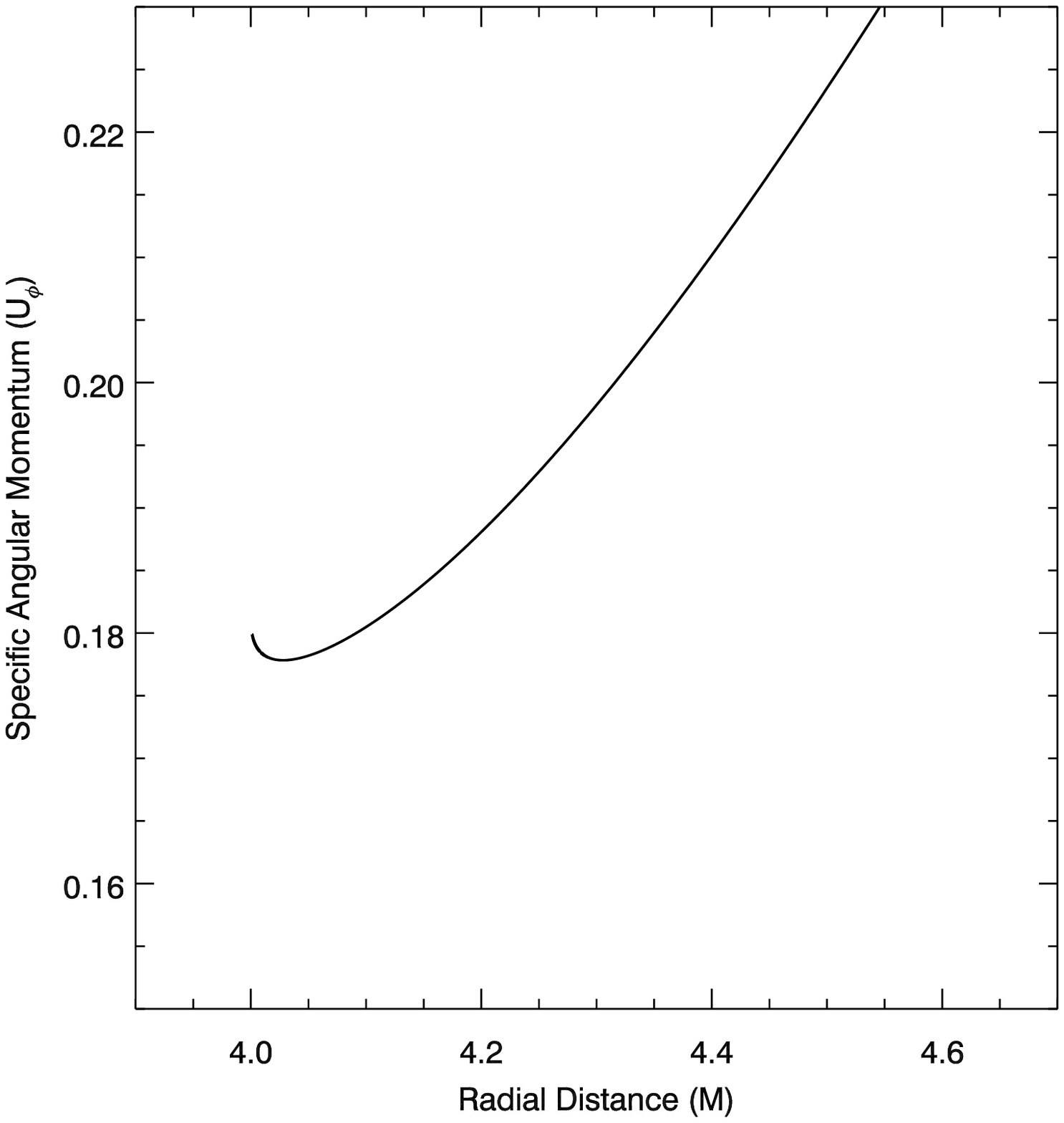}
\caption{Left panel shows the trajectory (dotted line) of the
(co-rotating) test particle just entering the suspension orbit just
outside the surface (solid circle) of the star with luminosity
$\left(\frac{L^{\infty}}{L^{\infty}_{Edd}}\right) \simeq 0.71$.
Right panel shows the profile of the specific angular momentum of the
test particle. The central star is spinning counter-clockwise
with an uniform angular momentum $j=0.1$.}
\end{figure}

It is rather puzzling that contrary to our expectation, the specific
angular momentum $U_{\phi}$ of the test particle instantly rises
instead of decreasing monotonically just before the test particle
enters the ``suspension orbit".  This behavior of the fine structure
can be interpreted as follows. And to this end, we refer to the
azimuthal component of the particle's geodesic equation given above
in equation (1). The right hand side of this geodesic equation
consists of three groups of terms.  The first group represents the
well-known Poynting-Robertson effect and among the four terms in
this group the second and the third terms are negligible because
they are multiplied by the radial component $U_{r}$ and the
azimuthal component $U_{\phi}$ which reduce to nearly zero as the
test particle enters the ``suspension orbit" (recall the defining
conditions for the ``suspension orbit" first given in \cite{OKL2010}
according to which $U_{r}$ drops to zero). As a result, the first
and the last terms dominate.  The second group appears to encode the
frame dragging effect of the central star as each term in this group
involves the Lense-Thirring angular velocity $\omega(r)$.  The
terms in this second group are small enough to be neglected as they
are multiplied by $U^{2}_{\phi}$ which is nearly zero as the test
particle enters the ``suspension orbit". Lastly, the third group
represents the radiation counter drag carefully studied in detail in
\cite{OKL2010}.  This group consists of three terms and each term
indicates different origin and hence different physical
interpretation.  To be more concrete, the first term is negligible
as it is multiplied by $U^{2}_{\phi}$ which is nearly zero as the
test particle enters the ``suspension orbit" and the third term is
nearly zero again as it is multiplied by $U_{r}$.  As a result,
the first and the third terms are both negligible.  The second
term which appears to have its origin in the frame dragging like the
two terms in the second group is non-vanishing and makes dominant
contribution as the test particle enters the ``suspension orbit". To
summarize, the terms on the right hand side of this azimuthal
component of the geodesic equation compete with one another
governing the behavior of the solution $U_{\phi}(r)$ and to be
more precise, the first and the fourth terms in the first group that
can be identified with the Poynting-Robertson effect or drag and the
second term in the third group which can be identified with the
radiation counter drag compete.  As a result of this competition
between two terms in the Poynting-Robertson effect represented by
the first group and the term in the radiation counter drag
represented by the third group, we end up with the puzzling behavior
of the test particle's specific angular momentum demonstrated by the
plot given above (Fig.1).  To be more concrete, it is puzzling as it
contradicts to the Kepler's law which dictates that as the test
particle falls into the central star, its angular velocity increases
monotonically whereas its angular momentum decreases monotonically.
We suspect that the advent of such puzzling behavior can be
attributed to the role played by the finite size effect of the
source as well as its luminosity and spin because if we neglect the
size of the source in the absence of the luminosity and spin, such
anomalous behavior would not happen in the first place.

\begin{figure}
\centering
\includegraphics[width=0.4\textwidth]{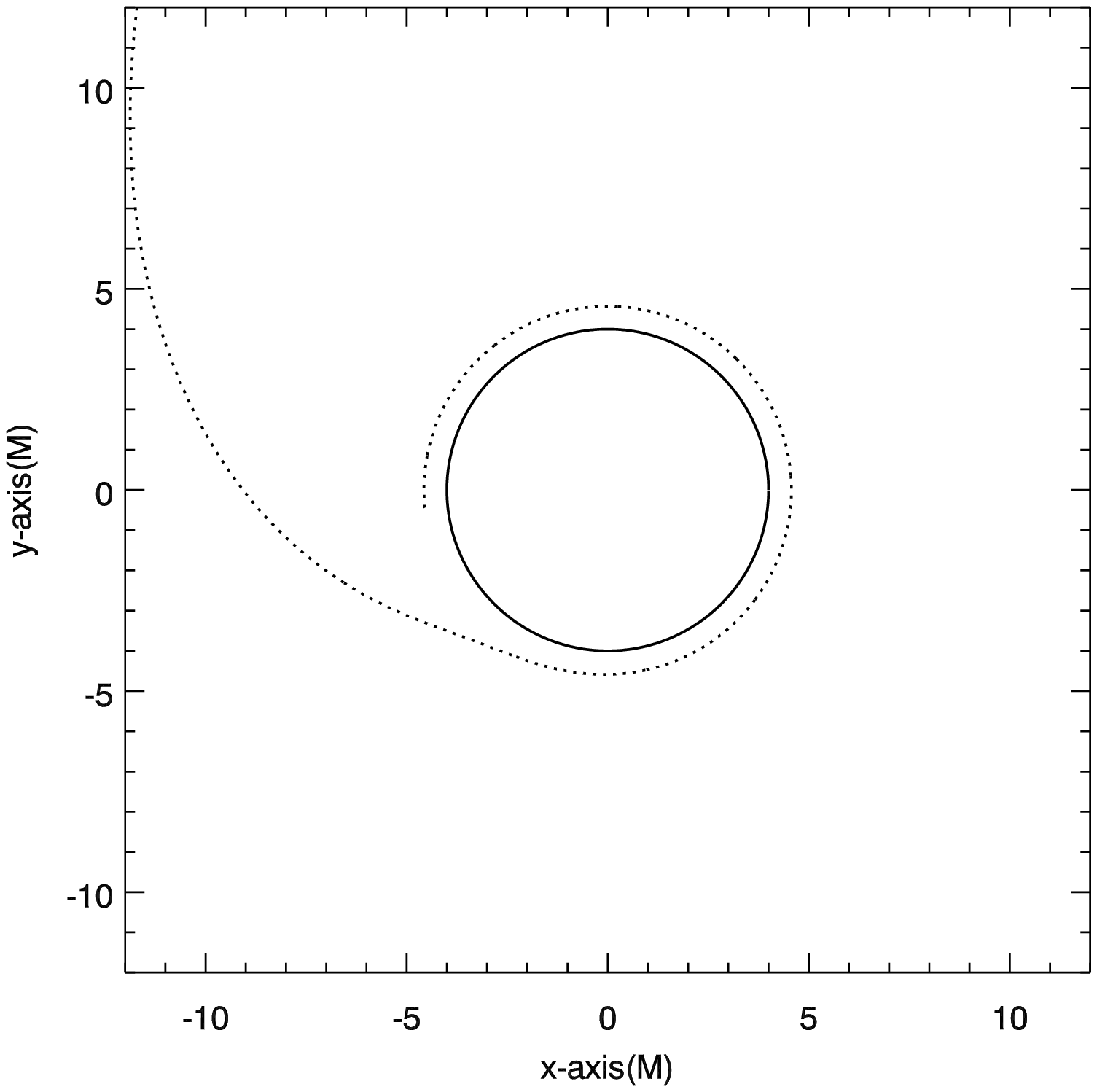}
\includegraphics[width=0.4\textwidth]{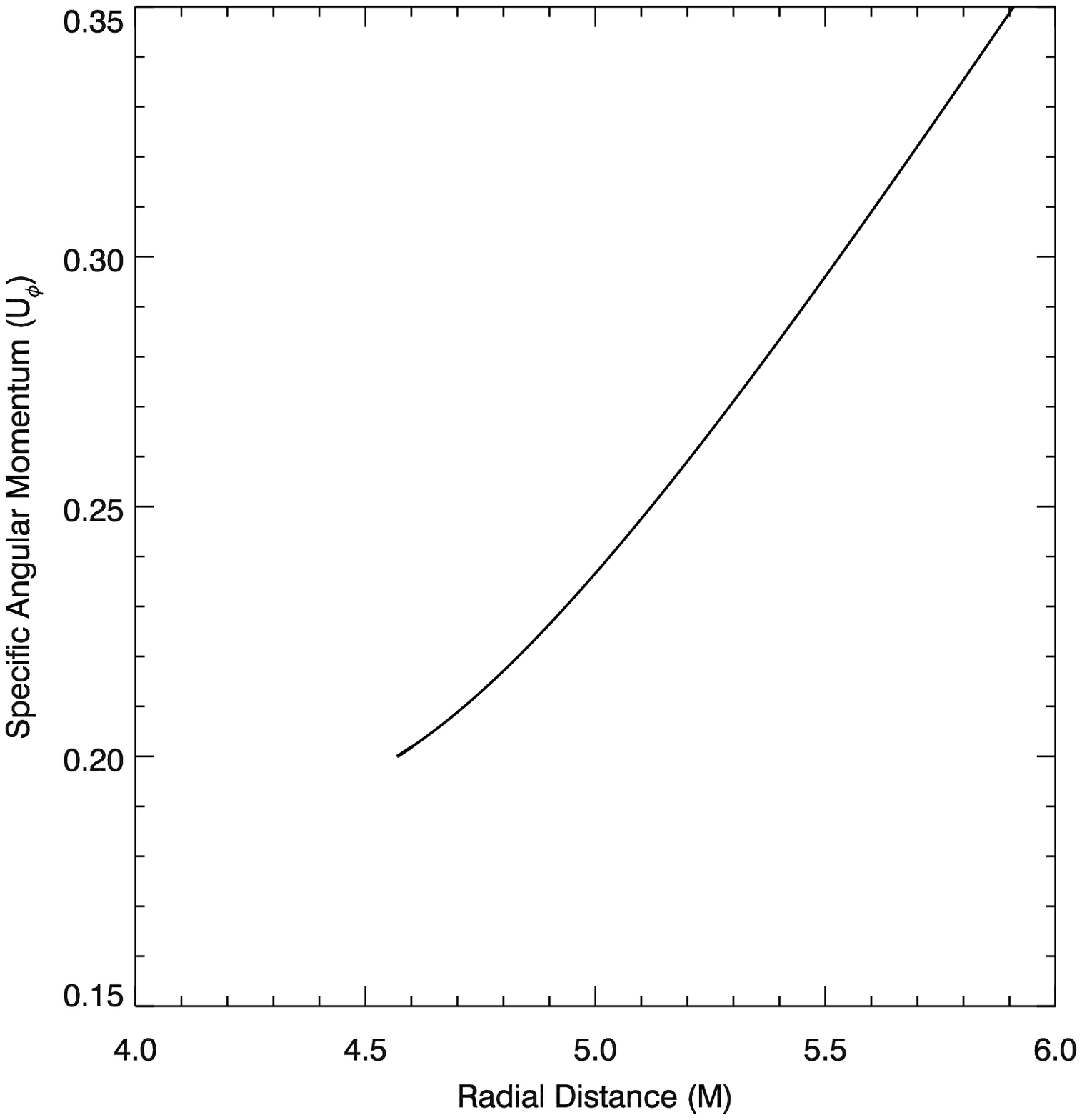}
\caption{Left panel shows the trajectory (dotted line) of the
(co-rotating) test particle just entering the suspension orbit
at some distance from the surface (solid circle) of the star
with luminosity
$\left(\frac{L^{\infty}}{L^{\infty}_{Edd}}\right) \simeq 0.75$.
Right panel shows the profile of the specific angular momentum of the
test particle. The central star is spinning counter-clockwise
with an uniform angular momentum $j=0.1$.}
\end{figure}

Now, we turn to the other case when the suspension orbit is at a
distance from the star's surface.  As can be seen in the plots
(Fig.2) of the numerical solution, the test particle's
specific angular momentum decreases monotonically as the test
particle falls into the ``suspension orbit" in consistent with our
general expectation based upon the Kepler's law stated above.  
This case, therefore, does not involve problematic issues that call for 
our careful investigation.

\begin{center}
{\rm\bf III. DISCUSSION}
\end{center}

To summarize, in the present work, as a sequel to our earlier work
(\cite{OKL2010}) where we studied the trajectory of test particle
near luminous rotating relativistic star, we explored the fine
structure of the trajectory of test particle just entering the
``suspension orbit" under the purpose of a detailed investigation of
test particle's trajectory in the vicinity of the ``suspension
orbit". Indeed, as a main result of our present work, we ended up
with a rather puzzling behavior that, contrary to our expectation,
the specific angular momentum $U_{\phi}$ of the test particle
instantly rises instead of decreasing monotonically just before the test
particle enters the ``suspension orbit".  We suspect that the advent
of such puzzling behavior can be attributed to the role played by
the finite size effect of the source as well as its luminosity and
spin because if we neglect the size of the source in the absence of
the luminosity and spin, such anomalous behavior would not happen in
the first place.  Lastly, therefore since the comprehensive physical interpretation
and the complete understanding of this anomalous behavior is not
available at the moment, we would like to pursue further
investigation along this line in our future works.

\vspace*{1cm}

\begin{center}
{\rm\bf Acknowledgements}
\end{center}

    This work was supported by the Korean Research Foundation Grant
No.2006-341-C00018. JSO acknowledges the support BK21 program to SNU.

\end{document}